\documentclass[reprint,showpacs,amsmath,amssymb,aps,prl]{revtex4-1}

\usepackage{epsf}
\usepackage{dcolumn}
\usepackage{bm}
\usepackage{hyperref}
\usepackage{color}

\begin{document}
\title{Adsorption and Pattern Recognition of Polymers at\\
Complex Surfaces with Attractive Stripe-like Motifs}
\author{Monika M\"oddel}
\email[E-mail: ]{Monika.Moeddel@itp.uni-leipzig.de}
\author{Wolfhard Janke}
\email[E-mail: ]{Wolfhard.Janke@itp.uni-leipzig.de}
\homepage[\\ Homepage: ]{http://www.physik.uni-leipzig.de/CQT.html}
\affiliation{Institut f\"ur Theoretische Physik and Centre for Theoretical Sciences (NTZ),
Universit\"at Leipzig, Postfach 100\,920, D-04009 Leipzig, Germany}
\author{Michael Bachmann}
\email[E-mail: ]{bachmann@smsyslab.org}
\homepage[\\ Homepage: ]{http://www.smsyslab.org}
\affiliation{Center for Simulational Physics, The University of Georgia,
Athens, GA 30602, USA}
\affiliation{Instituto de F\'{\i}sica,
Universidade Federal de Mato Grosso, Cuiab\'a (MT), Brazil}
\affiliation{Departamento de F\'{\i}sica, 
Universidade Federal de Minas Gerais, Belo Horizonte (MG), Brazil}

\begin{abstract}
		We construct the complete structural phase diagram of polymer
		adsorption at substrates with attractive stripe-like patterns in the
		parameter space spanned by the adsorption affinity of the stripes and
		temperature. Results were obtained by extensive
		generalized-ensemble Monte Carlo simulations of a generic model for the hybrid
		organic-inorganic system. By comparing with adhesion properties at
		homogeneous substrates, we find substantial differences in the formation of
		adsorbed polymer structures if translational invariance at the surface is
		broken by a regular pattern. Beside a more specific
		understanding of polymer adsorption processes, our results are
		potentially relevant for the design of macromolecular pattern recognition
		devices such as sensors. 
\end{abstract}
\maketitle

Interfacial macromolecular recognition is essential and ubiquitous in
biology processes and of potential interest for
nanotechnological applications. 
For these reasons, a thorough understanding of the generic features that
promote the adsorption of polymers at attractive substrates under the
influence of thermal fluctuations is of undeniable relevance. The
complex interplay of generic, undirected environmental effects (as, e.g.,
controlled by the temperature) and system-specific parameters that enable the
formation of stable, ordered structural phases of the polymer near the
substrate, requires systematic research. 

Consequently, the statistical mechanics of adsorption
transitions of entire classes of such hybrid systems can only be investigated
by means of efficient stochastic Monte Carlo computer simulations.
Computational studies have already been done extensively in the past for the
adsorption of lattice polymers and proteins
\cite{vrbova1,kumar1,bj2,prellberg2,bj3,binder2,allen1,ywli1}, and off-lattice
polymer models \cite{mbj1,liang1,mjb2} at homogeneous, flat substrates.
Various other geometries of substrates have been investigated as well, such as
polymer adsorption under confinement in spherical cavities~\cite{arkin1},
at cylindrical \cite{binderMilchev,srebnik1} and
fluctuating membrane-like surfaces \cite{skjb1}, and at nanowires
\cite{vb1,vmab1}. 
The recognition of substrates and surface patterns by polymers and proteins
has also been the subject of numerous experimental and computational studies
\cite{irbaeck1,bogner1,cerda1,khoklov1,straube1,bgbgij1,allen2}.
What is still lacking, but essential for the turn-over from empirical to
systematic design of macromolecular pattern-recognizing devices is the
understanding of the change of the generic structural behavior of
macromolecules in the vicinity of an attractive substrate, if the homogeneous
surface is replaced by a patterned interface. 

In this Letter, we compare the
structural phase diagrams of molecular adsorption at homogeneous and
heterogeneous substrates for entire classes of substrates that are
characterized by their adsorption propensity. We will unravel the complex
structure formation processes and the stable structural phases that are formed
by competing energetic interactions such as surface attraction strength and
intramolecular forces, and also entropic effects due to thermal activation and
the repercussions of finite-size effects. 

For our study, we introduce a generic model for the adsorption of a
self-interacting polymer at a complex surface with a stripe pattern. It is
sufficiently simple to enable a comprising computational study of all
structural phases of the hybrid system, but it is also specific
enough to identify the differences between polymer adsorption
behavior at homogeneous and patterned substrates.

The polymer is modeled by a linear bead-stick model with stiff bonds of length
unity. Non-bonded intramolecular interactions are described by a standard
Lennard-Jones potential; the sum over all pairwise contributions reads:
\begin{equation}
E_\text{LJ}/\varepsilon=4\sum_{i=1}^{N-2}\sum_{j=i+2}^{N}
\left[\left(\frac{\sigma}{r_{ij}}\right)^{12}-\left(\frac{\sigma}{r_{ij}}
\right)^{6}\right],
\end{equation}
where $r_{ij}$ is the distance between two non-bonded monomers $i$ and $j$;
$N=40$ is the number of monomers in the polymer chain. In our model,
the intramolecular interaction sets the overall energy scale $\varepsilon$,
in which also all other energies will be measured.
The length scale of this
interaction (van der Waals diameter) matches that of the bond length $r_{i\,
i+1}\equiv b$:
$\sigma/b=1$, which will serve as basic unit for all other lengths as well.
As a reference to DNA/RNA and protein systems, an effective
overall stiffness of the chain is introduced by the bending energy
$E_\text{bend}/\varepsilon=\epsilon_\text{bend}\sum_{i=1}^{N-2}
\left(1-\cos\vartheta_i\right),$
where $\vartheta_i$ is the bending angle between monomers $i$, $i+1$, and
$i+2$, and $\epsilon_\text{bend}=1/4$.
\begin{figure}
\centerline{\epsfxsize=8.6cm \epsfbox{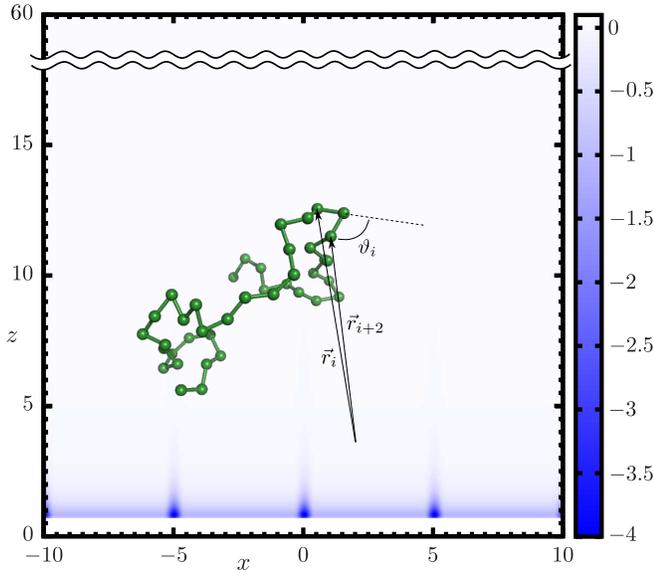}}
\caption{\label{fig:model} Polymer near a substrate with stripe pattern, with
the surface located at $z=0$. The density plot represents the periodic surface
potential. The steric wall at $z=60$ regularizes the
translational entropy in the half-space $z>0$. 
}
\end{figure}

The surface of the substrate is located at $z=0$ and possesses a
periodic stripe pattern that is oriented in $y$ direction. The bulk
of the substrate ($z<0$) is homogeneous. The interaction of 
the polymer chain with the patterned substrate is described
by~\cite{mbj1,mjb2}:
\begin{equation}
E_\text{s}/\varepsilon=\sum_{i=1}^{N}\epsilon_\text{sub}(x_i)
\left[\frac{2}{15}\left(\frac{\sigma_\text{s}}{z_i}\right)^{9}-
\left(\frac{\sigma_\text{s}}{z_i}
\right)^{3}\right]\quad (z_i>0),
\end{equation}
where the same length scale as above has been chosen
($\sigma_\text{s}/b=1$). The 9-3 Lennard-Jones-like potential follows
by
integrating a 12-6 Lennard-Jones potential over the half-space $z<0$. We
quantify the effect of the stripe pattern by the
periodic $x$-dependent dimensionless adsorption strength parameter:
\begin{equation}
\epsilon_\text{sub}(x)=\epsilon_\text{s}+\epsilon_\text{stripe}\left\{
\begin{array}{ll}
\cos^2(\alpha(x)\pi), & \mbox{if}\: |\alpha(x)|\le 1/2,\\
0, & \mbox{otherwise},
\end{array}
\right.
\end{equation}
where the choice $\alpha(x)=[(x/\sigma_x+1/2) \mod D] - 1/2$ \cite{rem1}
guarantees that the periodic potential is maximally attractive at the stripe
locations $x_\text{max}^{(k)}=\pm kD\sigma_x$ ($k$ integer), smoothly decays
towards $x_\text{max}^{(k)}\pm \sigma_x/2$, and is zero otherwise. As
for all other length scales, we set $\sigma_x/b=1$ in the
simulations. The
distance between the stripes was chosen to be $D=5$. 

Thus, the total energy of any polymer conformation is given by
$E=E_\text{LJ}+E_\text{bend}+E_\text{s}$.
The hybrid model and the effective surface potential
strength that is felt by each monomer are depicted in Fig.~\ref{fig:model}.
To prevent the non-grafted polymer from escaping, a steric wall is placed at
$z=60$. The influence of this constraint upon the translational entropy is
well understood~\cite{mjb2}. There are no boundaries in $x$ and $y$
directions.
\begin{figure}
\centerline{\epsfxsize=8.6cm \epsfbox{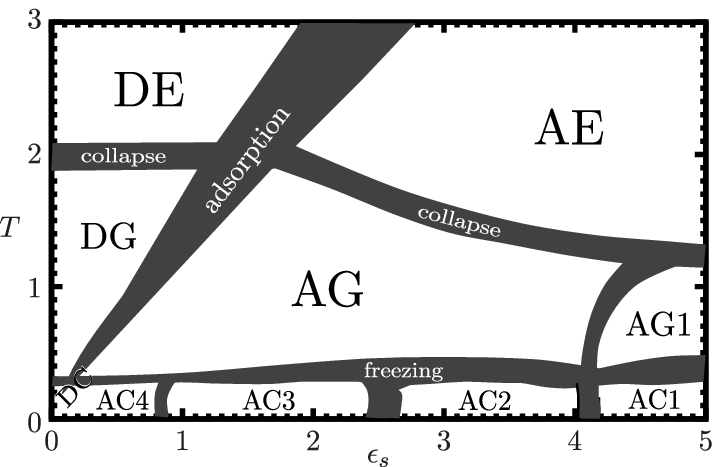}}

\centerline{\epsfxsize=8.6cm \epsfbox{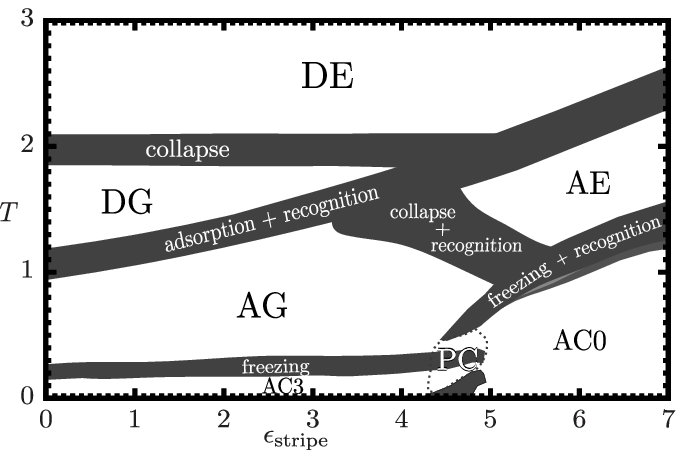}}
\caption{\label{fig:phasediagrams} Hyper-phase diagrams of structural polymer
phases for homogeneous substrates (top) with $\epsilon_{\rm stripe}=0$ and
for substrates with stripe pattern (bottom), where $\epsilon_s=1$. In
A/D phases polymer structures are preferably adsorbed/desorbed. The
second letter indicates increasing order in these phase regions:
expanded (E), globular (G), and compact (C);
PC is short for ``phase coexistence''. Temperatures $T$ are given in
units of $\varepsilon/k_\mathrm{B}$.}
\end{figure}

With this model, the two cases of substrate types, homogeneous and patterned,
can be compared systematically. Homogeneous substrates are represented by
$\epsilon_\text{stripe}=0$ and the only energy scale associated with the
surface potential that competes with the polymer
parameters, is governed by $\epsilon_\text{s}$. This case has been
investigated in detail
in Refs.~\cite{mbj1,mjb2}. The hyper-phase diagram for a 40-mer, parametrized
by $\epsilon_\text{s}$ and temperature $T$, is shown in
Fig.~\ref{fig:phasediagrams}(top). The more interesting case of the
patterned
substrate is simulated by adjusting the homogeneous component of the
surface energy by setting $\epsilon_\text{s}=1$ and by
considering
$\epsilon_\text{stripe}$ as a variable parameter. 

We simulated this system at 71 fixed values of
$\epsilon_\text{stripe}$ with the parallel tempering Monte Carlo
method~\cite{huku1}, using in each case 72 replicas at different temperatures.
The total number of sweeps 
amounted to $10^{10}$. 

The structural hyper-phase
diagram that corresponds to this case is shown in
Fig.~\ref{fig:phasediagrams}(bottom), also for a polymer with $N=40$ monomers.
This is the central result of our study. The
monotonic behaviors of various canonical response quantities, such
as the specific heat and fluctuations of structural quantities (gyration
tensor components, contact numbers), were investigated and regions of thermal
activity identified (peaks and ``shoulders''). The accumulation of these
signals is represented by the transition bands shown in the phase diagrams.
Since the system is finite, the width of the bands is a systematic
uncertainty~\cite{mbj1}. 
As usual, it should be noted that
(pseudo)transitions between
structural phases shall not be confused with
thermodynamic phase transitions, although the
origin of the structural transitions, cooperativity, is similar. In
exemplary simulations of longer chains with up to 80 monomers, we do not
observe qualitative changes in the phase behavior, i.e., the results for
$N=40$ are representative.

\begin{figure}
\centerline{\epsfxsize=8.6cm \epsfbox{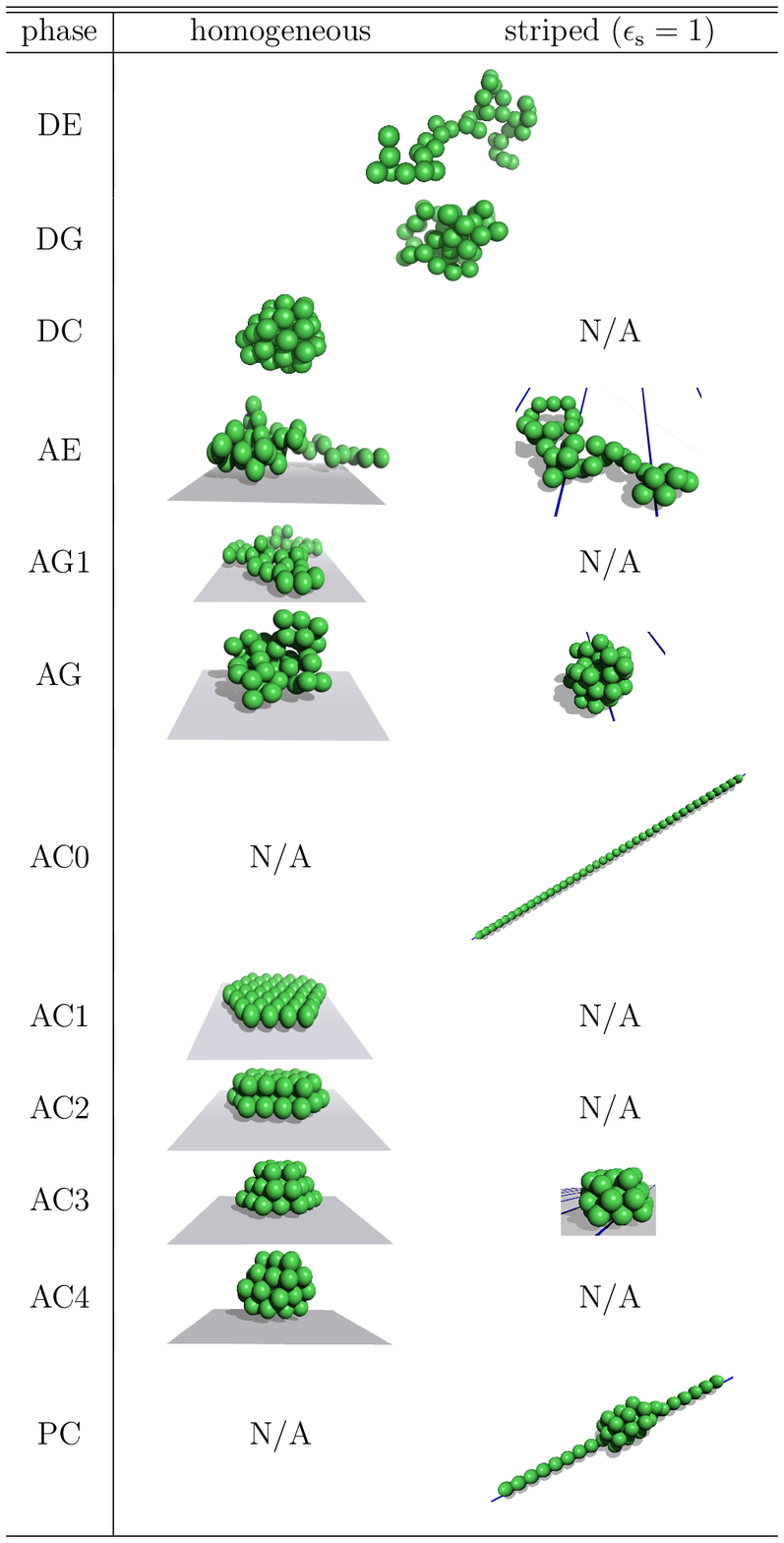}}
\caption{\label{fig:legend} Examples of typical conformations in the different
structural phases for homogeneous ($\epsilon_\text{stripe}=0$) and striped
substrates (with $\epsilon_\text{s}=1$).}
\end{figure}

Before discussing the novel features of adsorption behavior under the
influence of the pattern potential, let us review the main structural phases
of the homogeneous case first, as shown in Fig.~\ref{fig:phasediagrams}(top).
The adsorption transition line separates the major adsorbed phases of
expanded (random-coil) structures (AE), globular adsorbats (AG), and
compact, crystalline structures (AC) from the well-known desorbed phases of
expanded (DE), globular (DG), and compact conformations (DC). Particularly
noteworthy are the topological transitions from three-dimensional
conformations to two-dimensional films (AE$\rightarrow$AG1 and
AG$\rightarrow$AG1), as well as the layering transitions towards AC$n$, where
$n$ denotes the number of layers in the conformation~\cite{mbj1,mjb2}. By
comparing the results for various system sizes and also with lattice results,
no obvious indications could be found that the general phase structure will
substantially change towards the thermodynamic limit. Even the hierarchical
solid-solid (layering) transitions from mono- to multiple-layer phases are
surprisingly persistent. Representative polymer conformations are shown for
all phases in Fig.~\ref{fig:legend}. 

The structural phase diagram for polymer adsorption at a striped substrate,
as shown in Fig.~\ref{fig:phasediagrams}(bottom), is also generic in large
parts, but it depends on the overall attraction strength of the substrate (in
our model $\epsilon_\text{s}$). We chose $\epsilon_\text{s}=1$, because in
this case the energy scale of the overall attraction strength of the
substrate is identical to the non-bonded intramolecular energy scale. This
has the advantage that the limit $\epsilon_\text{stripe}=0$ is non-trivial
(standard adsorption at a homogeneous substrate) and the phase
structure in the chosen temperature interval is balanced, i.e., there are
distinct desorbed and adsorbed phases. Thus, the cross-sections at
$\epsilon_\text{stripe}=0$ in the phase diagram in
Fig.~\ref{fig:phasediagrams}(bottom) and at $\epsilon_\text{s}=1$ in 
Fig.~\ref{fig:phasediagrams}(top) are identical, which is why in the
desorbed region only DG and DE are present, but not DC. The most compact
polymer structures are those with three layers (AC3). The adsorption
transition line separates the globular regimes DG and AG. The general phase
structure and thus also the dominant polymer conformations are virtually the
same for both classes of substrates.
It should be noted that the correspondence between both substrate classes
holds also for other values of $\epsilon_\text{s}$, as long as
$\epsilon_\text{stripe}$ does not exceed a specific threshold value (here it
is $\epsilon_\text{stripe}\approx 3$, where phase AE starts mixing in). 

Regarding the polymer structures, the essential difference between the
adsorption processes on both types of substrates is that the polymer prefers
the contact to the energetically more favorable stripe regions on the
patterned substrate. Since the extension (radius of gyration in the $xy$
plane) of the compact conformations in AG and AC3 is smaller than the distance
between the stripes, the polymer recognizes exactly one stripe upon
adsorption. We here generally consider an adsorption transition to be a
recognition process, if the polymer adjusts to the surface pattern in any
form.
In this case, the space between the
stripes is virtually emptied, i.e., the AG/AC3 phases in the case of the
patterned substrate have a different appearance than their analogs in the
homogeneous case. Effectively, the presence of the stripes reduces the
translational entropy on the substrate. For the same reason, the adsorption
from DG to AG is a docking process with no apparent refolding. The increased
attraction affinity of the patterned substrate caused by the stripes leads to
an increased adsorption temperature. The freezing transition from AG to AC3
remains virtually unaffected by an increase of $\epsilon_\text{stripe}$ in
this region ($\epsilon_\text{stripe}<4.5$): Once the polymer has docked in
phase AG, it only reorders monomers upon further cooling to optimize the
number of internal contacts and the distance of each monomer to the stripe it
binds to.

In phase DE, entropy clearly dominates over non-bonded polymer energy and
conformations are unstructured. Lowering the temperature leads to adsorption,
but not ordering, i.e., the adsorption phase AE forms. The energetic
attraction of the stripes is larger than the homogeneous regions of the
substrate, so the polymer recognizes the existence of the stripes, but its 
typical extension is larger than the distance between two stripes. Therefore,
the polymer structures attach to several stripes simultaneously, but in no
specific way. For comparatively large stripe attraction strength
($\epsilon_\text{stripe}>6$), the polymer undergoes a direct transition from
AE to a singular regime that has no relevance on homogeneous substrates. This
is the ``rodlike phase'' AC0 of linelike structures, where all monomers prefer
contact with a single stripe (see Fig.~\ref{fig:legend}). 

It is a truly essential feature of stripe-patterned adsorption that with AC0
we have identified another topological phase. Remember that the 40mer has
four AC phases on homogeneous substrates, of which AC1 is filmlike, i.e.,
two-dimensional, whereas polymer structures in AC2, AC3, and AC4 form
the three-dimensional topological class of compact phases, where structures
extend into the third dimension perpendicular to the substrate. However, AC0
is apparently one-dimensional. Topological transitions between these phases
are supposed to be particularly strong and persistent in the thermodynamic
limit~\cite{bj2,prellberg2,mbj1,mjb2}.

Another remarkable feature is the transition from AC0 to AC3 by passing a
transition regime that we denote by PC (phase coexistence). Given the fact
that we have chosen thin stripes with orientational interaction directed
almost entirely into the direction perpendicular to the substrate, lamellar or
film-like double-rod structures (which would make up a phase AC1) and
double-layer or triple-rod structures (that would form a phase AC2) have to
compete with ``pearl-necklace'' structures as shown in Fig.~\ref{fig:legend}.
The result is the mixed phase PC, where the mentioned geometries coexist, but
none dominates. Mixed solid phases occur for finite polymers also in
DC~\cite{gvnb1}. 

To summarize, we have identified the complete phase structure of polymers
adsorbing at a substrate with stripe pattern by means of
parallel tempering
Monte Carlo simulations. By comparison with known results obtained
for homogeneous substrates, we found substantial differences in the
adsorption behavior, where the attractive interaction of the
stripes governs the formation of polymer structures at the adsorbent. We also
found that a directional stripe potential favors the formation of crystalline
droplets and rodlike strings at a single stripe. Consequently, the adsorption
transition in the globular regime (DE$\rightarrow$AE) and the
collapse/reordering transitions at the substrate (AE$\rightarrow$AG;
AG,AE$\rightarrow$AC0) were identified as the only transitions, where the
polymer recognizes the stripe pattern. To conclude, our general results
contribute to the systematic understanding of polymer adsorption and
recognition at patterned complex surfaces, which is relevant for non-empirical
approaches to the design of nanosensoric applications.

This work has been supported partially by the DFG (German Science 
Foundation) under Grant No.\ JA 483/24-3, by the NSF under
Grant No.\ DMR-1207437, and by CNPq (National
Council for Scientific and Technological Development, Brazil) 
under Grant No.\ 402091/2012-4.
Support by the Leipzig 
Graduate School of Excellence ``BuildMoNa'' and by 
the DFH-UFA (Franco--German University) is also acknowledged.


\begin{thebibliography}{10}
\bibitem{rem1}
The modulo operation $v \mod w$, where $v,w$ are real numbers, is given by the
floor function: $v \mod w=v-w\left\lfloor\frac{v}{w}\right\rfloor$, where
$\lfloor a \rfloor$ is the largest integer not greater than $a$.
\end{thebibliography}
\end{document}